\documentclass[conference]{IEEEtran}
\IEEEoverridecommandlockouts
\usepackage{cite}
\usepackage{amsmath,amssymb,amsfonts}
\usepackage{graphicx}
\usepackage{textcomp}
\usepackage{xcolor}
\usepackage{bm}
\usepackage{subfigure}
\usepackage{algorithm}
\usepackage{algorithmicx}
\usepackage{algpseudocode}
\renewcommand{\algorithmicensure}{\textbf{Output:}}

\def\BibTeX{{\rm B\kern-.05em{\sc i\kern-.025em b}\kern-.08em
    T\kern-.1667em\lower.7ex\hbox{E}\kern-.125emX}}
\begin{document}

\graphicspath{{../fig/}}

\title{Over-the-Air Gaussian Process Regression\\Based on Product of Experts}

\author{\IEEEauthorblockN{Koya Sato}
\IEEEauthorblockA{Artificial Intelligence eXploration Research Center,\\
The University of Electro-Communications, 1-5-1, Chofugaoka, Chofu-shi, Tokyo, Japan \\
E-mail: k\_sato@ieee.org}
\thanks{This work was supported in part by JST, ACT-X, JPMJAX21AA and JST SICORP, JPMJSC20C1.}
}

\maketitle

\begin{abstract}
    This paper proposes a distributed Gaussian process regression (GPR) with over-the-air computation, termed {\it AirComp GPR}, for communication- and computation-efficient data analysis over wireless networks. GPR is a non-parametric regression method that can model the target flexibly. However, its computational complexity and communication efficiency tend to be significant as the number of data increases. AirComp GPR focuses on that product-of-experts-based GPR approximates the exact GPR by a sum of values reported from distributed nodes. We introduce AirComp for the training and prediction steps to allow the nodes to transmit their local computation results simultaneously; the communication strategies are presented, including distributed training based on perfect and statistical channel state information cases. Applying to a radio map construction task, we demonstrate that AirComp GPR speeds up the computation time while maintaining the communication cost in training constant regardless of the numbers of data and nodes.
\end{abstract}


\section{Introduction}
\label{sect:intro}
Gaussian process regression (GPR) is a non-parametric approach to regression tasks, which realizes flexible modeling of a dataset without specifying low-level assumptions\,\cite{Rasmussen2004,pmlr-v37-deisenroth15}.
Assuming GP for the target data, we can obtain both the mean and variance of the regression results.
There has been a wide range of applications for GPR such as environmental monitoring based on spatial statistics\,\cite{gelfand2010handbook, Cressie}, experimental design\,\cite{srinivas-arxiv2009} and motion trajectory analysis\,\cite{6126365}; in wireless communication systems, recent results have shown its advances in coverage analysis and communication design, with the term of {\it radio map}\,\cite{bi-wirelesscommun2019, tobias-arxiv2022,sato-tccn2017}.
GPR will play an important role in the next Internet of Things (IoT) era.
\par
However, GPR has some critical drawbacks regarding communication and computational costs in such applications.
Let us consider a situation where multiple nodes are distributed on a network to monitor an environmental state and connected to a server wirelessly, as envisioned in \cite{s20113113}.
When the server performs GPR to analyze the sensing results, the nodes need to upload their sensing data to the server.
The exact GPR requires inverse matrices in training and prediction steps. This leads the complexity of $\mathcal{O}(N^3)$ for $N$ training data; further, for $n_\mathrm{in}$ input dimension data, the nodes upload $(n_\mathrm{in}+1)N$ variables to the server.
The first problem can be improved by distributed GPR based on the product of experts\,\cite{ng-arxiv2014, pmlr-v37-deisenroth15}.
This method approximates GPR by the sum of computation results at nodes to reduce the computational complexity from $\mathcal{O}(N^3)$ at the server to $\mathcal{O}((N/M)^3)$ at $M$ distributed nodes; however, the communication slots still depend on $M$.
\par
In this paper, toward a communication- and computation-efficient IoT monitoring system, we propose a distributed GPR scheme with over-the-air computation, termed {\it AirComp GPR}.
Over-the-air computation is a technique for communication-efficient distributed computation over shared channels based on nomographic functions\,\cite{mario-twc2015, yang-twc2020}.
Each node transmits its message with an analog modulation function. Then, the receiver obtains the target computation result from the superimposed signal based on a decoding function. Since multiple nodes transmit their analog-modulated signals simultaneously, we can realize a low-latency computation over networks.
We focus on that both training/regression results in the distributed GPR are based on the sum of computation results reported from the nodes. The proposed method aggregates the local computation results based on the over-the-air computation; as a result, the communication cost does not depend on the data size and the number of nodes.
\par
Major contributions of this paper are listed as follows.
\begin{itemize}
    \item We propose AirComp-aided distributed GPR for communication/computation efficient regression over wireless networks. It is shown that the computational complexity can be reduced from $\mathcal{O}(N^3)$ at BS to $\mathcal{O}((N/M)^3)$ at $M$ distributed nodes, and its communication cost at the training step can be constant regardless of $M$ and $N$.
    \item Two schemes are introduced for the training step: perfect channel state information (CSI)-based and statistical CSI-based schemes. The first approach can perform the distributed GPR with a limited accuracy degradation from full GPR; further, the latter enables no requirements for the uplink instantaneous channel estimations.
    \item Performance of AirComp GPR is analyzed in the radio map construction task. We demonstrate that an accurate radio map can be constructed efficiently.
\end{itemize}
\par
{\it Notations:} throughout this paper, the transpose, determinant, and inverse operators are denoted by $(\cdot)^\mathrm{T}, \mathrm{det}(\cdot)$ and $(\cdot)^{-1}$, while the expectation and the variance are expressed by $\mathbb{E}[\cdot]$ and $\mathrm{Var}[\cdot]$, respectively. Further, $|\cdot|$ and $||\cdot||$ are defined as operators to obtain the absolute and Euclidean distance, respectively.
\section{System Model}
\label{sect:systemmodel}
\subsection{Task Definition}
\label{subsec:task}
We consider a situation where $M$ sensing nodes are connected to a base station (BS) over wireless networks.
The $i$-th node has a dataset,
\begin{equation}
    \mathcal{D}_i = \left\{({\bm x}_{i,k}, y_{i,k})\mid k=1,2,\cdots,N_i\right\},
\end{equation}
where $N_i$ is the number of data, ${\bm x}_{i,k}$ is the input vector (e.g., sensing location) and $y_{i,k}=f({\bm x}_{i,k})+\epsilon$ is its output value generated from $\mathcal{N}(f({\bm x}_{i,k}), \sigma^2_\epsilon)$ ($\epsilon \sim \mathcal{N}(0, \sigma^2_\epsilon)$ is the independently and identically distributed (i.i.d.) noise).
When local datasets are non-overlapped each other, the full dataset over the network can be expressed as
\begin{equation}
    \mathcal{D} = \bigcup_{i=1}^{M}\mathcal{D}_i,
\end{equation}
where the number of full data can be defined as $N=\sum_{i=1}^{M}N_i$.
Further, it is assumed that all data in $\mathcal{D}$ follows a Gaussian process: i.e., $f\sim \mathrm{GP}\left(\mu({\bm x}), k({\bm x}, {\bm x}')\right)$, where $\mu({\bm x})$ is the expectation value at ${\bm x}$, and $k({\bm x}, {\bm x}')$ is the covariance between $\mu({\bm x})$ and $\mu({\bm x}')$.
The task in this context is to estimate $f$ for test inputs ${\bm X}_\ast = \left[{\bm x}_{\ast,1},{\bm x}_{\ast,2},\cdots, {\bm x}_{\ast,n_\mathrm{test}}\right]$ from $\mathcal{D}_i$ distributedly.
\par
Possible applications of the above task include environmental monitoring\,\cite{6385653} and radio map construction\,\cite{bi-wirelesscommun2019}.

\subsection{Signal Model}
\label{subsec:signalmodel}
\begin{figure}[t]
    \centering
    \includegraphics[width=1.0\linewidth]{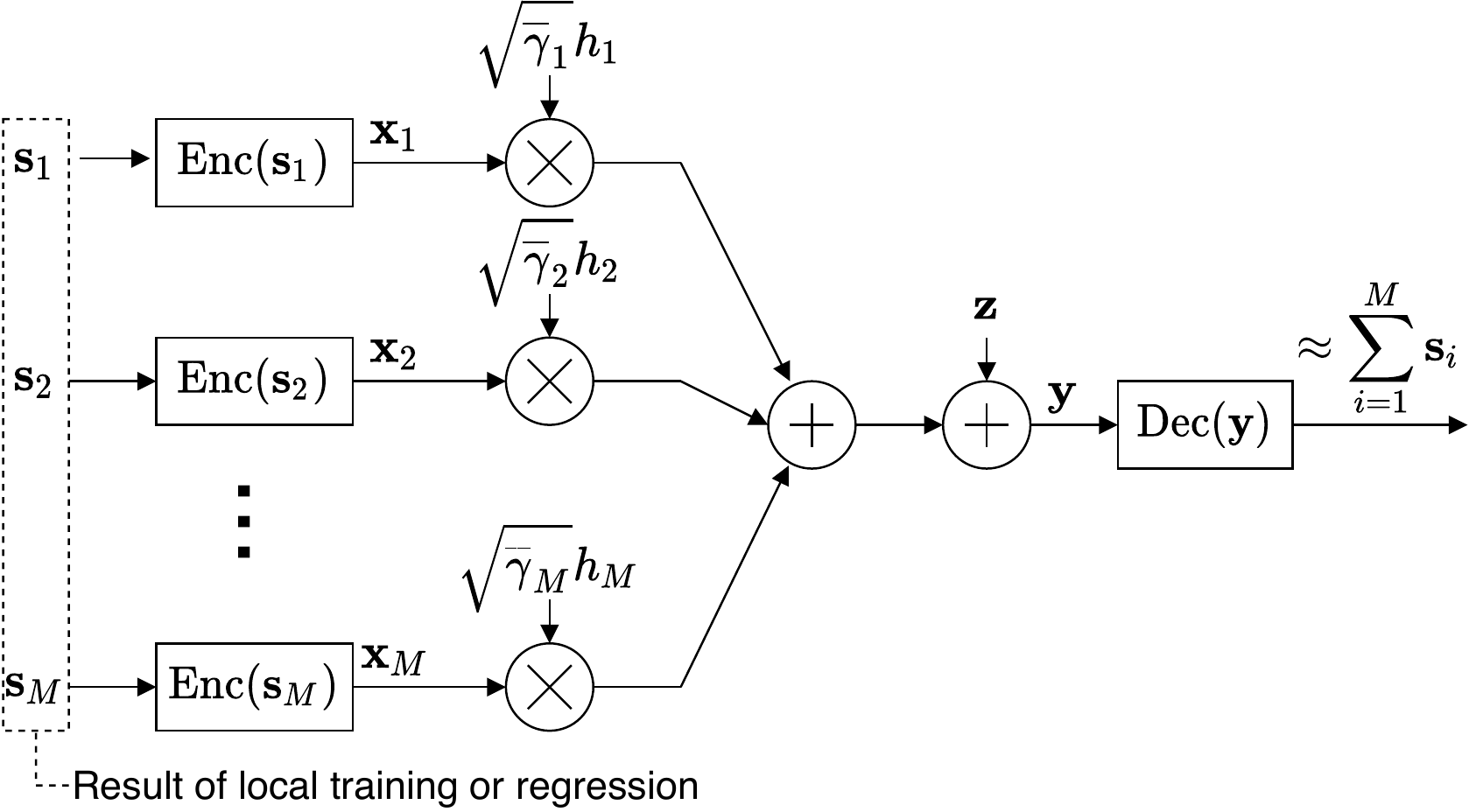}
    \caption{Signal transmission model.}
    \label{fig:signalmodel}
\end{figure}
AirComp GPR can be divided into training and regression steps.
We herein define the signal model for these steps.
Fig.\,\ref{fig:signalmodel} summarizes the signal transmission model, where all nodes simultaneously transmit their messages to BS through a shared wireless channel.
The $i$-th node first encodes its message $\mathbf{s}_i$ so that BS can extract the sum of $\mathbf{s}_i$; we denote this process as $\mathbf{x}_i = \mathrm{Enc}(\mathbf{s}_i)$.
When all nodes are time synchronized, the received signal at BS can be given by
\begin{equation}
    \mathbf{y} = \sum_{i=1}^{M}\sqrt{\overline{\gamma_i}}h_i \mathbf{x}_i + \mathbf{z},
\end{equation}
where $\sqrt{\overline{\gamma}_i} \in \mathbb{R}$ is the average channel gain and $h_i \sim \mathcal{CN}(0, 1)$ is the i.i.d.\,instantaneous channel gain assuming flat over one transmission. Further, $\mathbf{x}_i$ is the transmitted vector constrained by the maximum transmission power $P_\mathrm{max}$ as $||\mathbf{x}_i||^2\leq P_\mathrm{max}$, and $\mathbf{z}$ is the additive white Gaussian noise (AWGN) vector following $\mathcal{CN}(0, \sigma_z^2)$, where $\sigma_z^2$ is the noise floor.
Then, BS extract the sum of $\mathbf{s}_i$ using a decoding operation, defined by $\mathrm{Dec}(\mathbf{y})$.
\par
Note that BS has to share a message that contains a few hyper-parameters in the training step to the nodes. To enable the channel estimation at the nodes, BS broadcasts it with digital encoding with sufficient transmission power; we assume that the nodes can decode it correctly.
Further, assuming channel reciprocity, the nodes can estimate the instantaneous channel state information (CSI) $\sqrt{\overline{\gamma_i}}h_i$ owing to the broadcasted downlink signals.
In contrast, we consider two situations for BS; (a) global CSI and (b) statistical CSI (i.e., only $\overline{\gamma_i}$ is available). This condition affects the AirComp in the training step (see \ref{subsec:proposed_training}).
% measurement model

% over-the-air signal model
\section{Gaussian Process Regression}
Before explaining the proposed method, this section introduces full GPR, and its distributed method based on products of experts\,\cite{pmlr-v37-deisenroth15}.
Note that, for simplicity, this section assumes $n_\mathrm{test}=1$ and denotes the test input as ${\bm x}_\ast$.
 \subsection{Full GPR}
 \label{subsec:fullgpr}
 Consider a situation where BS has the full dataset $\mathcal{D}$ and performs the exact GPR.
 From the full dataset $\mathcal{D}$, we define ${\bm y}=\{y_{i,k} \mid \forall{(i,k)}\}$ and ${\bm X}=\{{\bm x}_{i,k}\mid \forall{(i,k)}\}$.
 GPR first needs to tune hyper-parameters ${\bm \theta} = \left\{{\bm \psi}, \sigma_\epsilon \right\}$, where ${\bm \psi}$ is the hyper-parameter vector for a kernel function $k$. Finding ${\bm \theta}$ can be realized by maximizing the log-marginal likelihood,
 \begin{align}
 \log p\left({\bm y}| {\bm X}, {\bm \theta}\right) = &-\frac{1}{2}({\bm y}-{\bm m})^\mathrm{T}\left({\bm K} + \sigma^2_\epsilon {\bm I}\right)^{-1}({\bm y}-{\bm m}) \nonumber\\
& - \frac{1}{2}\log\mathrm{det}\left({\bm K} + \sigma^2_\epsilon {\bm I}\right) -\frac{N}{2}\log2\pi,
  \label{eq:log-marginal}
 \end{align}
 where ${\bm I}$ is the $N \times N$ identity matrix and ${\bm K}\in \mathbb{R}^{N \times N}$ is the kernel matrix, where its element is $K_{ij} = k({\bm x}_i, {\bm x}_j)$ (${\bm x}_i$ is the $i$-th element in ${\bm X}$).
 Further, ${\bm m}$ is a vector with $N$ elements, where its $i$-th element $m({\bm x}_i)$ is the prior mean at ${\bm x}_i$\footnote{For example, vector ${\bm m}$ is given from $m({\bm x}_1)=m({\bm x}_2)=\cdots =m({\bm x}_N)=\frac{1}{N}\sum_{i=1}^{N}y_i$, where $y_i$ is the $i$-th element in ${\bm y}$.}.
 \par
  Based on a vector ${\bm \theta}$, the full GPR predicts the distribution of the output at the test input ${\bm x}_\ast$ as the Gaussian distribution with mean ($\mathbb{E}[f({\bm x}_\ast)] = \mu ({\bm x}_\ast)$) and variance ($\mathrm{Var}[f({\bm x}_\ast)]= \sigma^2 ({\bm x}_\ast)$) given by the following equations, respectively:
  \begin{align}
    \mu ({\bm x}_\ast) &=  m({\bm x}_\ast) + {\bm k}^{\mathrm{T}}_{\ast}\left({\bm K} + \sigma^2_\epsilon {\bm I}\right)^{-1}({\bm y}-{\bm m}) \label{eq:fullgpr-mean}\\
    \sigma^2 ({\bm x}_\ast) &= k_{\ast\ast} - {\bm k}^{\mathrm{T}}_{\ast}\left({\bm K} + \sigma^2_\epsilon {\bm I}\right)^{-1} {\bm k}_\ast, \label{eq:fullgpr-var}
  \end{align}
  where ${\bm k}_\ast = k({\bm X}, {\bm x}_\ast)$ and $k_{\ast\ast} = k({\bm x}_\ast, {\bm x}_\ast)$.
  \par
 %%computational complexity of full GPR
 A critical drawback of the full GPR is its computational complexity; that is, it requires $\mathcal{O}(N^3)$ to calculate the inverse of $N \times N$ matrices in Eqs.\,\eqref{eq:log-marginal}-\eqref{eq:fullgpr-var}.
 
 %% Product of Exparts
 \subsection{Distributed GPR Based on Products-of-GP-Experts}
 GPR can be parallelized based on the product of computations by distributed nodes termed {\it experts}; we refer to this method as {\it DGPR-PoEs} later.
 Similar to the model defined in \ref{subsec:task}, DGPR-PoEs divides the full dataset into subdatasets $\left\{\mathcal{D}_1, \mathcal{D}_2, \cdots, \mathcal{D}_M\right\}$ and distributes them to nodes.
 We detail the training and prediction steps below.
 \subsubsection{Training}
 \label{subsubsec:poe-training}
DGPR-PoEs assumes that local dataset $\mathcal{D}_i$ is independent of each other.
 Under this assumption, the marginal likelihood for the full dataset can be approximated by the product of the local values.
 This can be expressed as
 \begin{equation}
    p\left({\bm y}| {\bm X}, {\bm \theta}\right) \approx \prod_{i=1}^{M}p \left({\bm y}_i| {\bm X}_i, {\bm \theta}\right),
 \end{equation}
 where ${\bm y}_i=[y_{i, 1}, y_{i, 2}, \cdots, y_{i, N_i}]$, ${\bm X}_i=[{\bm x}_{i, 1}, {\bm x}_{i, 2}, \cdots, {\bm x}_{i, N_i}]$, and $p \left({\bm y}_i| {\bm X}_i, {\bm \theta}\right)$ is based on the $\mathcal{D}_i$.
 The hyper-parameter can be approximately found by maximizing its logarithmic form, i.e.,
 \begin{equation}
     \log p\left({\bm y}| {\bm X}, {\bm \theta}\right) \approx \sum_{i=1}^{M} \log p \left({\bm y}_i| {\bm X}_i, {\bm \theta}\right),
     \label{eq:log-marginal_approx}
 \end{equation}
where $\log p \left({\bm y}_i| {\bm X}_i, {\bm \theta}\right)$ is calculated at each node from
 \begin{align}
    \log p\left({\bm y}_i| {\bm X}_i, {\bm \theta}\right) &\!=\! 
    -\!\frac{1}{2}({\bm y}_i\!-\!{\bm m}_i)^\mathrm{T}\left({\bm K}_{\psi, i} \!+\! \sigma^2_\epsilon {\bm I}\right)^{-1}({\bm y}_i\!-\!{\bm m}_i)  \nonumber \\
    &-\!\frac{1}{2}\log\mathrm{det}\left({\bm K}_{\psi,i} + \sigma^2_\epsilon {\bm I}\right)-\frac{N_i}{2}\log2\pi. \label{eq:log-marginal_approx_local}
 \end{align}
 Note that ${\bm K}_{\psi,i} = k\left({\bm X}_i, {\bm X}_i\right)$ and ${\bm m}_i$ is the prior mean vector for ${\bm y}_i$ obtained from $\mathcal{D}_i$.
 For example, the hyper-parameter training based on the above approximation can be realized by iterating the following steps: (i)\,BS distributes ${\bm \theta}^{(t)}$, (ii)\,each node computes Eq.\,\eqref{eq:log-marginal_approx_local} based on its local dataset and uploads the result, and (iii)\,BS updates ${\bm \theta}^{(t)}$ to ${\bm \theta}^{(t+1)}$ based on an optimization algorithm (e.g., Nelder-Mead simplex\,\cite{neldermead-1965}).

 \subsubsection{Prediction}
 DGPR-PoEs estimates $f({\bm x}_\ast)$ based on
 \begin{equation}
   p(f({\bm x}_\ast)|{\bm x}_\ast, \mathcal{D}) = \prod_{i=1}^M p(f({\bm x}_\ast)|{\bm x}_\ast, \mathcal{D}_i).
 \end{equation}
 To perform the distributed GPR, each node estimates mean $\mu_i ({\bm x}_\ast)$ and variance $\sigma^2_i({\bm x}_\ast)$ based on Eqs.\,\eqref{eq:fullgpr-mean}\eqref{eq:fullgpr-var} and the local dataset.
 Then, mean and variance for $p(f({\bm x}_\ast)|{\bm x}_\ast, \mathcal{D})$ can be calculated by the following equations, respectively.
 \begin{align}
    \mu^\mathrm{poe}({\bm x}_\ast) &= \left(\sigma^\mathrm{poe}({\bm x}_\ast)\right)^2 \sum_{i=1}^{M}\left(\sigma_i({\bm x}_\ast)\right)^{-2} \mu_i({\bm x}_\ast), \label{eq:poes-mean} \\
    \left(\sigma^\mathrm{poe}({\bm x}_\ast)\right)^{-2}&= \sum_{i=1}^{M} \left(\sigma_i({\bm x}_\ast)\right)^{-2}.\label{eq:poes-var}
 \end{align}
 \par
 These training and prediction require each node to compute the inverse of $\left({\bm K}_{\psi, i} + \sigma^2_\epsilon {\bm I}\right)$. Thus, the computational complexity at the $i$-th node follows $\mathcal{O}\left((N/M)^3\right)$ when $N_1=N_2=\cdots=N_M$.

\section{AirComp GPR}
Let us apply DGPR-PoEs to the regression analysis over wireless networks.
DGPR-PoEs improves the computational complexity of GPR. However, 
digital transmissions will need $M$ slots to obtain a log-marginal likelihood value; further, computing mean and variance also requires $2Mn_\mathrm{test}$.
Here, as can be seen from Eqs.\,\eqref{eq:log-marginal_approx}\eqref{eq:poes-mean}\eqref{eq:poes-var}, both training and prediction are realized with the sum of reports from nodes.
Focusing on this feature, we propose an AirComp-enhanced DGPR-PoEs for communication/computation efficient regression analysis.

\subsection{Training}
\label{subsec:proposed_training}
\subsubsection{Perfect CSI-Based Method}
\label{subsubsec:proposed_perfectCSI}
As exemplified in \ref{subsubsec:poe-training}, finding an appropriate hyper-parameter requires to maximize Eq.\,\eqref{eq:log-marginal_approx} based on an iterative algorithm.
Thus, we consider an iterative training and denote the transmitted message, its encoded signal, and channel coefficient at the $t$-th step as $\mathbf{s}_{i}^{(t)}$, $\mathbf{x}^{(t)}_{i}$ and $h^{(t)}_i$, respectively.
\par
At the $t$-th step, BS first broadcasts ${\bm \theta}^{(t)}$ to the nodes.
%% encoding
The $i$-th node calculates the local log-marginal likelihood $L_i^{(t)}=\log p \left({\bm y}_i| {\bm X}_i, {\bm \theta}^{(t)}\right)$.
This node next encodes its message, $\mathbf{s}_{i}^{(t)} = \left[L_i^{(t)}\right]$, to a complex signal based on
\begin{align}
    \mathbf{x}^{(t)}_{i} = \mathrm{Enc}\left(\mathbf{s}_{i}^{(t)}\right)=\frac{\sqrt{\rho^{(t)}}}{\sqrt{\overline{\gamma}_i}h_i^{(t)}} \mathbf{s}_{i}^{(t)},
    \label{eq:encode_perfect}
\end{align}
where $\rho^{(t)}$ is a scalar for the power control at the slot $t$. BS determines this value as
\begin{align}
    \sqrt{\rho^{(t)}} = \min \left\{\frac{\sqrt{\overline{\gamma}_i}|h_i^{(t)}|\sqrt{P_\mathrm{max}}}{||\mathbf{s}_{i}^{(t)}||} \;\middle|\; i=1, 2, \cdots, M\right\}.
    \label{eq:powercontrol_perfect}
\end{align}
After the nodes transmit their messages simultaneously, BS receives the aggregated signal, which is derived by
\begin{equation}
  \mathbf{y}^{(t)} = \sqrt{\rho^{(t)}}\sum_{i=1}^{M} \left[L_{i}^{(t)}\right] + \left[z^{(t)}\right],
\end{equation}
where $z^{(t)}$ is the AWGN at the $t$-th slot.
%% decoding
Thus, BS can extract the sum of messages by taking the following operation:
\begin{equation}
    \mathrm{Dec}\left(\mathbf{y}^{(t)}\right) = \mathrm{Re}\left(\frac{\mathbf{y}^{(t)}}{\sqrt{\rho^{(t)}}}\right) \approx \log p({\bm y} | {\bm X}, {\bm \theta}) + \frac{z_\mathrm{R}^{(t)}}{\sqrt{\rho^{(t)}}},
    \label{eq:decoder-perfect}
\end{equation}
where $\mathrm{Re}(\cdot)$ is the operation to extract the real part, and $z_\mathrm{R}^{(t)}$ is the real part of the AWGN. For $\rho^{(t)}\rightarrow \infty$, we can compute $\mathrm{Dec}\left(\mathbf{y}^{(t)}\right) \approx \log p({\bm y} | {\bm X}, {\bm \theta})$.

\par
After the decoding, the BS updates ${\bm \theta}^{(t)}$ to ${\bm \theta}^{(t+1)}$ based on its optimizer, and iterates above process until Eq.\,\eqref{eq:decoder-perfect} is fully maximized.
This paper updates ${\bm \theta}^{(t)}$ based on multi-start local search\,\cite{Marti2018} with Nelder-Mead simplex\,\cite{neldermead-1965}.
The multi-start method iterates finding the local solution with differently initialized hyper-parameters to find a more good solution. Further, Nelder-Mead simplex is a heuristic optimizer to find the minimum of an objective function in a multidimensional space that can efficiently search for a local solution based on only the output of the objective function: a combination of the multi-start local search and Nelder-Mead simplex can efficiently tune the hyperparameters with avoiding local optima.
\par
We summarize this training algorithm, including the statistical CSI-based method, in Alg.\,\ref{alg:proposed_training}. Note that Alg.\,\ref{alg:proposed_training} defines the numbers of a training in Nelder-Mead simplex and multi-starts as $T$ and $T_\mathrm{multi}$, respectively.
Totally, this algorithm requires $T\times T_\mathrm{multi}$ iterations.

\begin{figure}[t]
  \begin{algorithm}[H]
    \caption{Distributed training algorithm for AirComp GPR based on multi-start Nelder-Mead simplex}
    \label{alg:proposed_training}
    \small
    \begin{algorithmic}
        \Require $P_\mathrm{max}$, $T$, $T_\mathrm{multi}$
        \Require $\{L_\mathrm{min}, L_\mathrm{max}\}$ (if statistical CSI case)
        \For{$t_\mathrm{multi}=0, \cdots, T_\mathrm{multi}-1$}
          \State{BS randomizes $\theta^{(0)}$.}
          \State{BS calculates $\rho^{(0)}$.}
          \For{$t=0, \cdots, T-1$}
            \If{perfect CSI case}
              \State{BS updates $\rho^{(t)}$ based on Eq.\,\eqref{eq:powercontrol_perfect}.}
            \EndIf
            \State{BS broadcasts $\theta^{(t)}$ and $\rho^{(t)}$.}
            \For{$i=1,2,\cdots,M$ \textbf{parallel}}
              \State{The $i$-th node calculates $L_i^{(t)}$ and encodes it into $\bf{x}$$_i^{(t)}$.}
              \State{The $i$-th node transmits $\bf{x}$$_i^{(t)}$ to BS.}
            \EndFor
            \State{BS decodes approximate of $\sum_{i=1}^{M}L_i^{(t)}$ from $\bf{y}$$^{(t)}$.}
            \State{BS updates $\theta^{(t)}$ to $\theta^{(t+1)}$ based on Nelder-Mead simplex.}
          \EndFor
        \EndFor
    \State \Return The best ${\bm \theta}^{(T-1)}$ over $T_\mathrm{multi}$ steps.

    \end{algorithmic}
  \end{algorithm}
\end{figure}

\subsubsection{Statistical CSI-Based Method}
\label{subsubsec:statistical}
Although the perfect CSI-based method can approximate the marginal-log likelihood well, it requires the BS to collect the global CSI $h_i^{(t)}$ and $||\mathbf{s}_{i}^{(t)}||$ every steps.
To improve this practical drawback, we also introduce a statistical CSI-based power control and training, having the following two features:
\begin{itemize}
    \item BS controls the transmission power at $t=0$ only, based on $\{\overline{\gamma}_i \mid i=1,2,\cdots,M\}$.
    \item At the encoding step, each node compensates the phase shift only, and does not consider the amplitude compensation.
\end{itemize}
Note that the instantaneous channel gain, $\sqrt{\overline{\gamma}_i}h_i^{(t)}$, is also available at the $i$-th node in this case owing to digitally-broadcasted sequences, that contains ${\bm \theta}^{(t)}$, from the BS.
\par
In this method, each node converts $L_i$ so that the maximum value of $||\mathbf{s}_i^{(t)}||$ is $\frac{1}{2} \left(L_\mathrm{max} + L_\mathrm{min}\right)$, where $\{L_\mathrm{min}, L_\mathrm{max}\}$ is a set of truncation parameters, with the following operation:
\begin{align}
    &\mathbf{s}_i^{(t)} \!=\!\! \left[\min\{\max\{L^{(t)}_i, L_\mathrm{min}\}, L_\mathrm{max}\}
    \!-\! \frac{1}{2}\left(L_\mathrm{max} \!+\! L_\mathrm{min}\right)\right].
\end{align}
Based on the above conversion, BS adjusts the uplink transmission power at $t=0$ so that the node with the lowest value of $\overline{\gamma}_i$ can transmit the signal with the maximum transmission power based on the following equation.
\begin{equation}
    \sqrt{\rho^{(0)}} = \min \left\{\frac{\sqrt{\overline{\gamma}_i}\sqrt{P_\mathrm{max}}}{\frac{1}{2}\left(L_\mathrm{max} + L_\mathrm{min}\right)} \;\middle|\; i=1, 2, \cdots, M\right\}.
    \label{eq:powercontrol_stat}
\end{equation}
All nodes follow $\rho^{(0)} = \rho^{(1)} = \cdots = \rho^{(T-1)}$ overall the training process.
\par
At the $t$-th step, the $i$-th node encodes $\mathbf{s}_i^{(t)}$ so that the phase shift by $h_i$ can be compensated: i.e.,
\begin{align}
    \mathbf{x}^{(t)}_{i} = \mathrm{Enc}\left(\mathbf{s}_{i}^{(t)}\right)=\sqrt{\rho^{(t)}} \frac{\overline{h}^{(t)}_i}{\sqrt{\overline{\gamma}_i}\left|h_i^{(t)}\right|} \mathbf{s}_{i}^{(t)},
    \label{eq:encode_stat}
\end{align}
where $\overline{h}^{(t)}_i$ is the conjugate of $h^{(t)}_i$.
Then, the aggregated signal at the BS can be derived by
\begin{equation}
  \mathbf{y}^{(t)}=\sqrt{\rho^{(t)}}\sum_{i=1}^{M}\left|h^{(t)}_i\right|\mathbf{s}_i^{(t)} + \mathbf{z}^{(t)}.
\end{equation}
Finally, with a similar operation shown in the perfect case (Eq.\,\eqref{eq:decoder-perfect}), BS decodes the sum of local log-marginal likelihood with the following computation
\begin{align}
    \mathrm{Dec}\left(\mathbf{y}^{(t)}\right) &= \mathrm{Re}\left(\frac{\mathbf{y}^{(t)}}{C\sqrt{\rho^{(t)}}}\right) \nonumber \\
    &=\frac{1}{C}\sum_{i=1}^{M}\left|h^{(t)}_i\right|L_i^{(t)} + \frac{z_\mathrm{R}^{(t)}}{C\sqrt{\rho^{(t)}}},
    \label{eq:decoder_stat}
\end{align}
where $C$ is a scalar designed to achieve an unbiased estimation for $\sum_{i=1}^{M}L_i^{(t)}$.
Here, $|h_i^{(t)}|$ is independent with $L_i^{(t)}$, and $\mathbb{E}[|h_1^{(t)}|]=\mathbb{E}{|h_2^{(t)}|}=\cdots=\mathbb{E}[|h|]$; thus, we can obtain the condition
\begin{equation}
  \mathbb{E}\left[\sum_{i=1}^{M}\left|h^{(t)}_i\right|L_i^{(t)}\right]=\mathbb{E}\left[|h|\right]\mathbb{E}\left[\sum_{i=1}^{M}L_i^{(t)}\right].
\end{equation}
For the Rayleigh fading channel, when $\sqrt{\rho^{(t)}}\rightarrow \infty$, an unbiased estimation can be realized with
\begin{equation}
  C = \mathbb{E}\left[|h|\right] =\frac{\sqrt{\pi}}{2}.
\end{equation}
Note that $C=1$ for AWGN channels.

\subsection{Prediction}
Next, the nodes perform the distributed regression based on the fully-trained hyper-parameter vector ${\bm \theta}_\mathrm{opt}$.
The number of test points $n_\mathrm{test}$ tends to be sufficiently large to analyze the function $f$ over a wide range of test inputs in practice: for example, a radio map construction over 500-m $\times$ 500-m area with ten-meter grids requires $n_\mathrm{test}=2500$.
This feature implies that the overhead required for the channel estimation between the nodes and BS can be sufficiently small.
Considering the above, this prediction step considers the perfect CSI case only; however, it can be extended to the statistical CSI case as with \ref{subsubsec:statistical}.
\par
According to Eqs.\,\eqref{eq:poes-mean}\eqref{eq:poes-var}, the distributed GPR can be realized based on (i) the sum of $(\sigma_i({\bm x}_\ast))^{-2}$ and (ii) $(\sigma_i({\bm x}_\ast))^{-2}\mu_i({\bm x}_\ast)$. Because both $\sigma_i({\bm x}_\ast)$ and $\mu_i({\bm x}_\ast)$ are calculated at the nodes locally, we can apply AirComp to this prediction step, as with the training step.
\par
The nodes first calculate mean and variance for the test inputs based on  Eqs.\,\eqref{eq:poes-mean}\eqref{eq:poes-var}. We denote these results at the $i$-th node as the two vectors.
\begin{align}
    {\bm \mu}^\mathrm{poe}_{i} &= \left[\mu^\mathrm{poe}_i({\bm x}_{\ast,1}), \mu^\mathrm{poe}_i({\bm x}_{\ast,2}), \cdots, \mu^\mathrm{poe}_i({\bm x}_{\ast,n_\mathrm{test}})\right], \\
    {\bm \sigma}^\mathrm{poe}_{i} &= \left[\sigma^\mathrm{poe}_i({\bm x}_{\ast,1}), \sigma^\mathrm{poe}_i({\bm x}_{\ast,2}), \cdots, \sigma^\mathrm{poe}_i({\bm x}_{\ast,n_\mathrm{test}})\right].
\end{align}
Then, the $i$-th node generates the following two signal vectors.
\begin{itemize}
    \item $\mathbf{s}_i^{(0)}$: a vector with $n_\mathrm{test}$ elements, representing its $j$-th element as $(\sigma^\mathrm{poe}_i({\bm x}_{\ast,j}))^{-2}$.
    \item $\mathbf{s}_i^{(1)}$: a vector with $n_\mathrm{test}$ elements, representing its $j$-th element as $(\sigma^\mathrm{poe}_i({\bm x}_{\ast,j}))^{-2}\mu^\mathrm{poe}_i({\bm x}_{\ast,j})$.
\end{itemize}
Over two communication slots, the nodes transmit $\mathbf{s}_i^{(0)}$ and $\mathbf{s}_i^{(1)}$ respectively, to BS based on AirComp presented in \ref{subsubsec:proposed_perfectCSI}.
Finally, BS can obtain the regression results, $\mu^\mathrm{poe}({\bm x}_{\ast, i})$ and $\sigma^\mathrm{poe}({\bm x}_{\ast, i})$, from $\sum_{i=1}^{M}\mathbf{s}_i^{(0)}$ and $\sum_{i=1}^{M}\mathbf{s}_i^{(1)}$ with Eqs.\,\eqref{eq:poes-mean}\eqref{eq:poes-var}.

\subsection{Regression Example}
\begin{figure}[t]
    \centering
    \subfigure[Regression behavior.]{\includegraphics[width=0.48\linewidth]{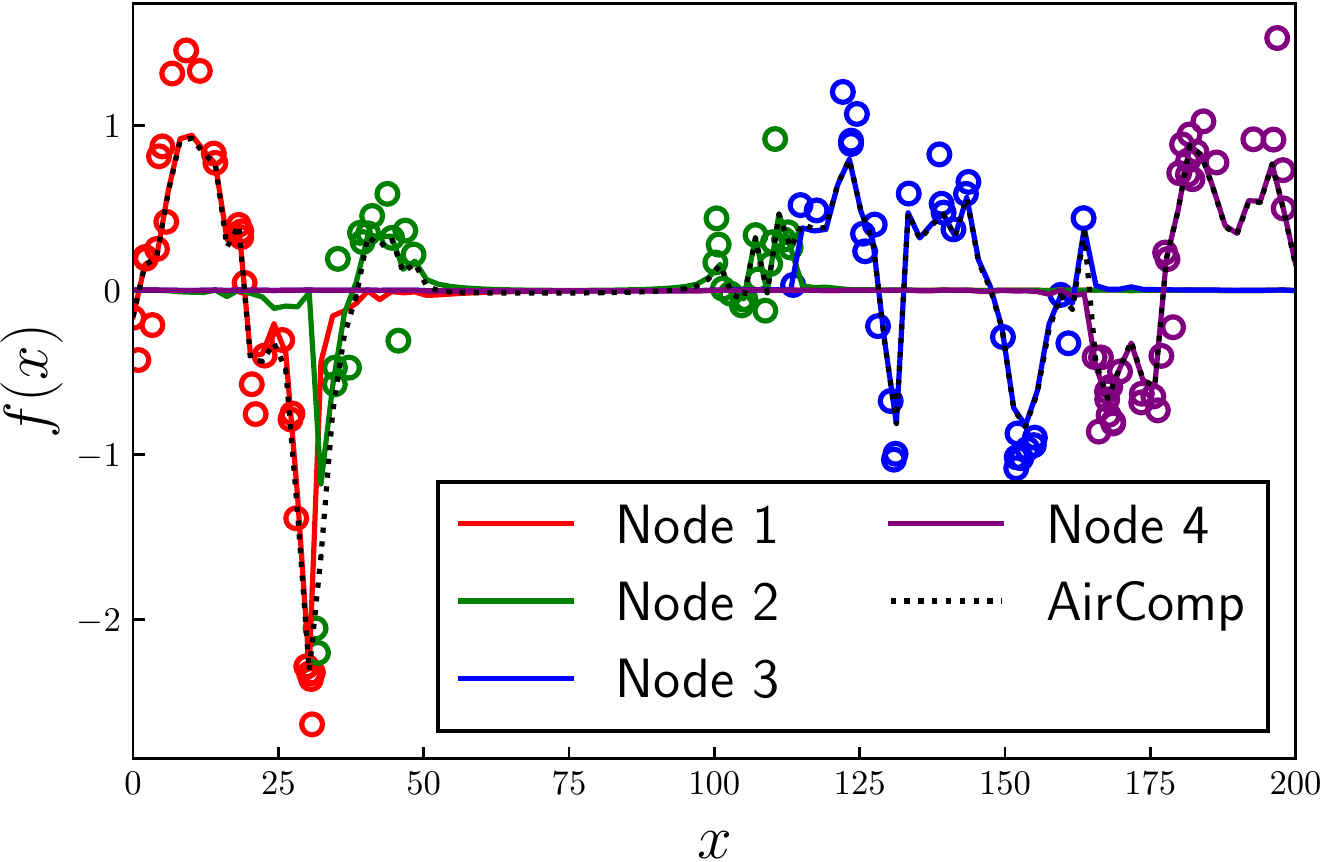}
    \label{subfig:behavior}}
    \subfigure[Estimated range of 95-percentile.]{\includegraphics[width=0.48\linewidth]{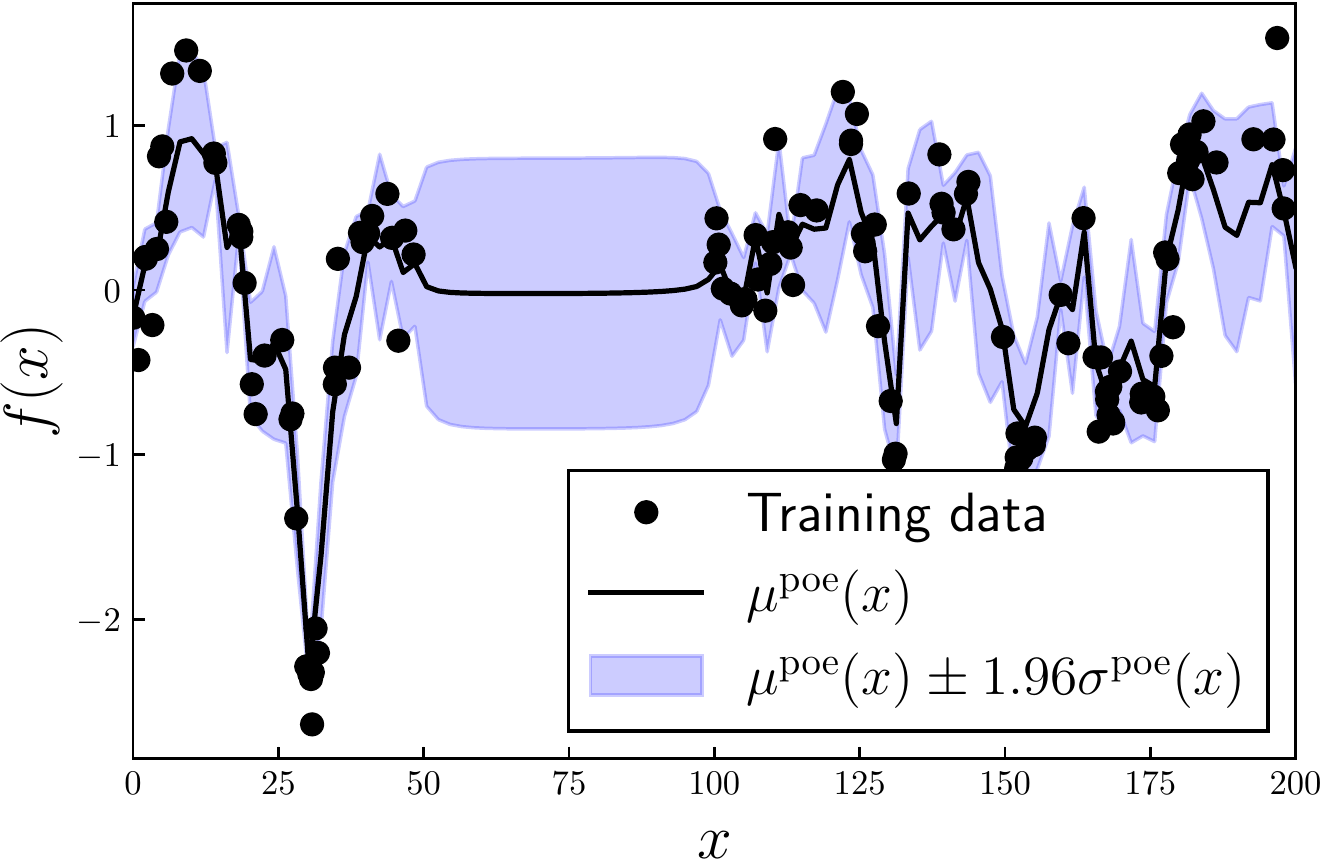}
    \label{subfig:example}}
  \caption{Example of AirComp GPR ($M=4, N=128$ and $\overline{\gamma}_i=1$).}\label{fig:example_proposedgpr}
\end{figure}

Fig.\,\ref{fig:example_proposedgpr} demonstrates a regression example based on AirComp GPR. This dataset is generated from a pure GP with zero-mean and the variance is 1. As the kernel $k$, we set an exponential kernel defined by
\begin{equation}
  k({\bm x}_i, {\bm x}_j| {\bm \psi}) = \psi_1 \exp\left(-\frac{||{\bm x}_i - {\bm x}_j||}{\psi_2}\right),
  \label{eq:kernel}
\end{equation}
where ${\bm \psi} = [\psi_1, \psi_2]$ and $\psi_1, \psi_2 > 0$.
Distributed training data, regression result at nodes, and the AirComp result are showon in Fig.\,\ref{subfig:behavior}. We plot the local computation result at the $i$-th node, $(\sigma^\mathrm{poe}_i({\bm x}_{\ast}))^{-2}\mu^\mathrm{poe}_i({\bm x}_{\ast})$, scaled by $(\sigma^\mathrm{poe}({\bm x}_\ast))^2$; the AirComp result indicates the sum of local computation results.
This figure demonstrates that each node estimates detailed fluctuation near its local training data, and near mean values are output elsewhere.
\par
Fig.\,\ref{subfig:example} plots $\mu^\mathrm{poe}\pm 1.96 \sigma^\mathrm{poe}$; i.e., the estimated 95-percentile. It can be seen that the unobserved region shows high uncertainties.

\section{Performance in Radio Map Construction}
\label{sect:performance}
This section presents the performances of AirComp GPR under a radio map construction task.
Radio map visualizes the spatial distribution of the received signal power values\,\cite{bi-wirelesscommun2019,sato-tccn2017}. Since the typical received signal power follows GP over the dB domain, the full GPR can obtain an optimal radio map from $\mathcal{D}$. We present how AirComp GPR works in this task\footnote{
Several works applied Kriging to radio map (or radio environment map) construction tasks\,\cite{5711699,sato-tccn2017}. Note that Kriging is equivalent to GPR in this case\,\cite{Rasmussen2004}.}.

\subsection{Simulation Setup}
\begin{table}[t]
    \caption{Simulation Parameters}
    \label{table:parameter}
    \centering
    \begin{tabular}{l|r}\hline
        Path loss index $\eta$ & 3 \\ \hline
        Transmission power at target transmitter $P_\mathrm{Tx}$ & 10\,[dBm] \\ \hline
        Shadowing standard deviation $\sigma_\mathrm{dB}$ & 8\,[dB] \\ \hline
        Correlation distance $d_\mathrm{cor}$ & 100\,[m] \\ \hline
        Number of workers $M$ & 4 \\ \hline
        Number of data $N$ & 128 \\ \hline
        Number of local data $N_i$ & $N/M$ \\ \hline
        Maximum transmission power $P_\mathrm{max}$ & 10\,[dBm] \\ \hline
        Average channel gain $\overline{\gamma}_i$ & -50\,[dB] \\ \hline
        Noise floor $\sigma_z^2$ & -90\,[dBm] \\ \hline
        Numbers of iterations $T$ and $T_\mathrm{multi}$ & 600 and 3 \\ \hline
        Truncation parameters $\{L_\mathrm{min}, L_\mathrm{max}\}$ & $\{-5000, 0\}$ \\ \hline
        Kernel function & Eq.\,\eqref{eq:kernel} \\ \hline
    \end{tabular}
\end{table}

This simulation constructs a radio map (i.e., the spatial distribution of average received signal power values) for a transmitter over one-dimensional space.
Assuming the transmitter is located at a two-dimensional coordinate ${\bm x}_\mathrm{Tx}=[0, 500\,\mathrm{[m]}]$, the $i$-th node measures $N/M$ received signal power values over ${\bm x}_{i,k} = [l_{i,k}, 0]\;\forall k$; ${\bm x}_{i,k}$ denotes the $k$-th measurement location by the $i$-th node, and its $x$ coordinate is randomly selected from $1\,\text{[m]}\leq l_{i,k} \leq 1000\,\text{[m]}$.
When a node at a location measures a fully-averaged received signal power, the received signal power can be expressed as
\begin{equation}
    P_\mathrm{Rx}({\bm x}_{i,k}) = P_\mathrm{Tx} - 10\eta \mathrm{log}_{10}||{\bm x}_\mathrm{Tx} - {\bm x}_{i,k}|| + W\,\mathrm{[dBm]},
\end{equation}
where $P_\mathrm{Tx}$ is the transmission power, $W$ is a shadowing that follows spatially correlated normal distribution with zero mean and standard deviation $\sigma_\mathrm{dB}$. The two shadowing values at ${\bm x}_i$ and ${\bm x}_j$ are correlated based on the exponential decay model\,\cite{Gudmundson-el1991}. This equation is modeled as
\begin{equation}
    \mathrm{Cor}\left[W({\bm x}_i), W({\bm x}_j)\right] = \mathrm{exp}\left(-\frac{||{\bm x}_i - {\bm x}_j||}{d_\mathrm{cor}}\ln 2\right),
\end{equation}
where $d_\mathrm{cor}$ is the correlation distance.
\par
Each node constructs its local dataset by
\begin{equation}
    \mathcal{D}_i = \{(d_{i,k}, P_\mathrm{Rx}(d_{i,k}))\mid k=1,2,\cdots, \lfloor N/M \rfloor\},
\end{equation}
and computes the mean vector ${\bm m}_i$ based on ordinary least squares (OLS). A root mean squared error (RMSE) is evaluated at test locations, selected from unobserved regions, assuming $n_\mathrm{test}=10$.
After the above evaluation is iterated 1000 times, we calculate the mean of RMSEs and use it as the evaluation result.
\par
In addition to AirComp GPR, we evaluate the following methods: (i) full GPR, (ii) ideal DGPR-PoEs, and (iii) perfect path loss estimation.
In (i), BS performs the full GPR based on the full dataset; further, method (ii) performs DGPR-PoEs without any communication error and noise to discuss how much the analog modulation part in AirComp GPR affects accuracy.
Finally, the method (iii) has $P_\mathrm{Tx}$ and $\eta$ as prior. However, it cannot estimate $W$.
Simulation parameters follow Table\,\ref{table:parameter}, unless otherwise noted\footnote{Assuming the maximum of $T$ iterations, we finished one Nelder-Mead simplex operation when the updated amount of the objective function from the previous iteration is less than $10^{-4}$.}.

% \subsection{Communication and Computation Performances}
\subsection{Evaluation Results}
\label{subsec:evaluation-results}
\begin{figure}[t]
    \centering
    \includegraphics[width=1.00\linewidth]{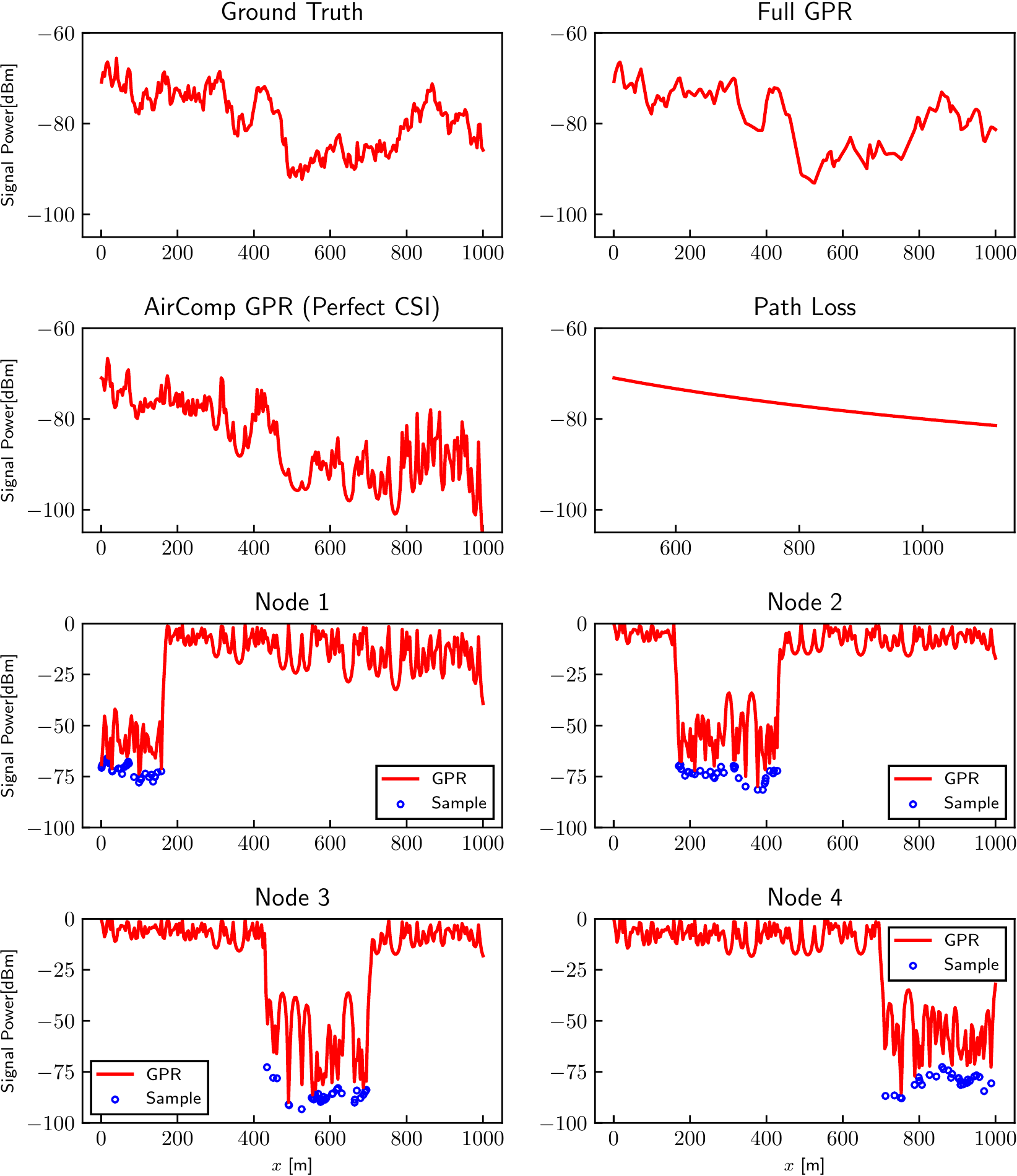}
    \vspace{-5mm}
  \caption{Radio map construction example (one-dimensional case where $M=4$ and $N=128$).}
  \label{fig:example-radiomap}
\end{figure}

We first show a radio map construction example in Fig.\,\ref{fig:example-radiomap}.
To show the relationship between measurement data and GPR results, we divided the measurement area into $M$ regions and assigned them to each node. Further, each regression result indicates $(\sigma^\mathrm{poe}_i({\bm x}_{\ast}))^{-2}\mu^\mathrm{poe}_i({\bm x}_{\ast})$ scaled by $(\sigma^\mathrm{poe}({\bm x}_\ast))^2$.
The local prediction results tend to be zero in the area away from the measurement points; this trend implies that local computation influences AirComp results in the vicinity of the measurement points and suppresses its influence in other areas.
After the AirComp, the aggregated computation results can estimate the trend of shadowing across the entire area.
\par
\begin{figure}[t]
    \centering
    \includegraphics[width=0.9\linewidth]{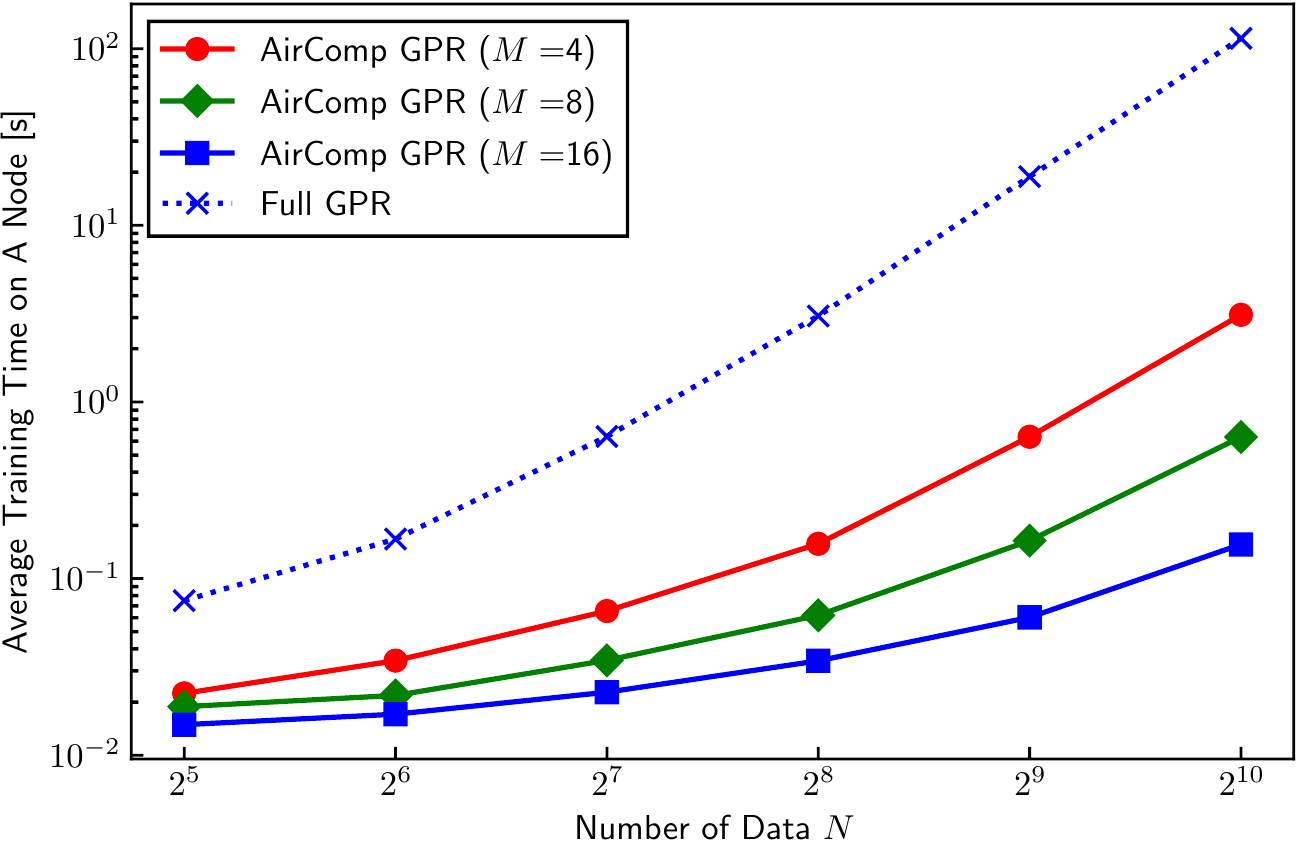}
    % \vspace{-5mm}
  \caption{Effects of $N$ on training time.}
  \label{fig:comp-time}
%   \label{fig:effects-nodes}
\end{figure}
Fig.\,\ref{fig:comp-time} indicates effects of $N$ on the average training time performance.
We implemented this simulation code based on Python 3.9.12 and numpy 1.21.5 and ran it on AMD Ryzen 5950X with DDR4-2133 128GB memory; single physical core was enabled to evaluate the training time at a node.
In AirComp GPR, the nodes perform their local training in parallel. Thus, this performance expresses the time spent over the network in a training step.
AirComp GPR can reduce the training time compared with full GPR in various conditions, and this improvement becomes significant as $M$ increases (e.g., 733x faster than full GPR when $N=2^{10}$ and $M=16$).
\par
\begin{table}[t]
    \caption{Equivalent Number of Variables Transmitted in Uplink}
    \label{table:comparison_variables}
    \centering
    \begin{tabular}{l|c|c}\hline
        Method      & Analog or Digital & Number of Variables   \\ \hline \hline
        Full GPR    & Digital           & $(n_\mathrm{in}+1)N$  \\
        Ideal DGPR-PoEs    & Digital           & $M(T\cdot T_\mathrm{multi}+2n_\mathrm{test})$  \\
        AirComp GPR & Analog            & $T\cdot T_\mathrm{multi}+2n_\mathrm{test}$   \\ \hline
    \end{tabular}
\end{table}

Table\,\ref{table:comparison_variables} shows equivalent numbers of variables transmitted in uplink over a pair of training and regression steps.
We count an equivalent number as one when multiple nodes transmit their local variables simultaneously.
Full GPR requires the nodes to upload their local datasets to BS with digital transmission. Thus, it requires $(n_\mathrm{in}+1)N$ ($n_\mathrm{in}$ is the number of dimensions in the input vector).
In contrast, ideal DGPR-PoEs needs $M(T\cdot T_\mathrm{multi}+2n_\mathrm{test})$ variables to collect $M(T\cdot T_\mathrm{multi})$ local likelihood values in the training step and $2Mn_\mathrm{test}$ regression results, including both mean and variance, in the regression step.
Finally, AirComp GPR takes $1/M$ smaller than ideal DGPR-PoEs owing to $M$ simultaneous transmissions.
For example, when $N=1024$, full GPR requires 2048 (3072 for 2D radio maps). Further, ideal DGPR-PoEs and AirComp GPR require $1820M$ and 1820, respectively.
\par
\begin{figure}[t]
    \centering
    \includegraphics[width=0.9\linewidth]{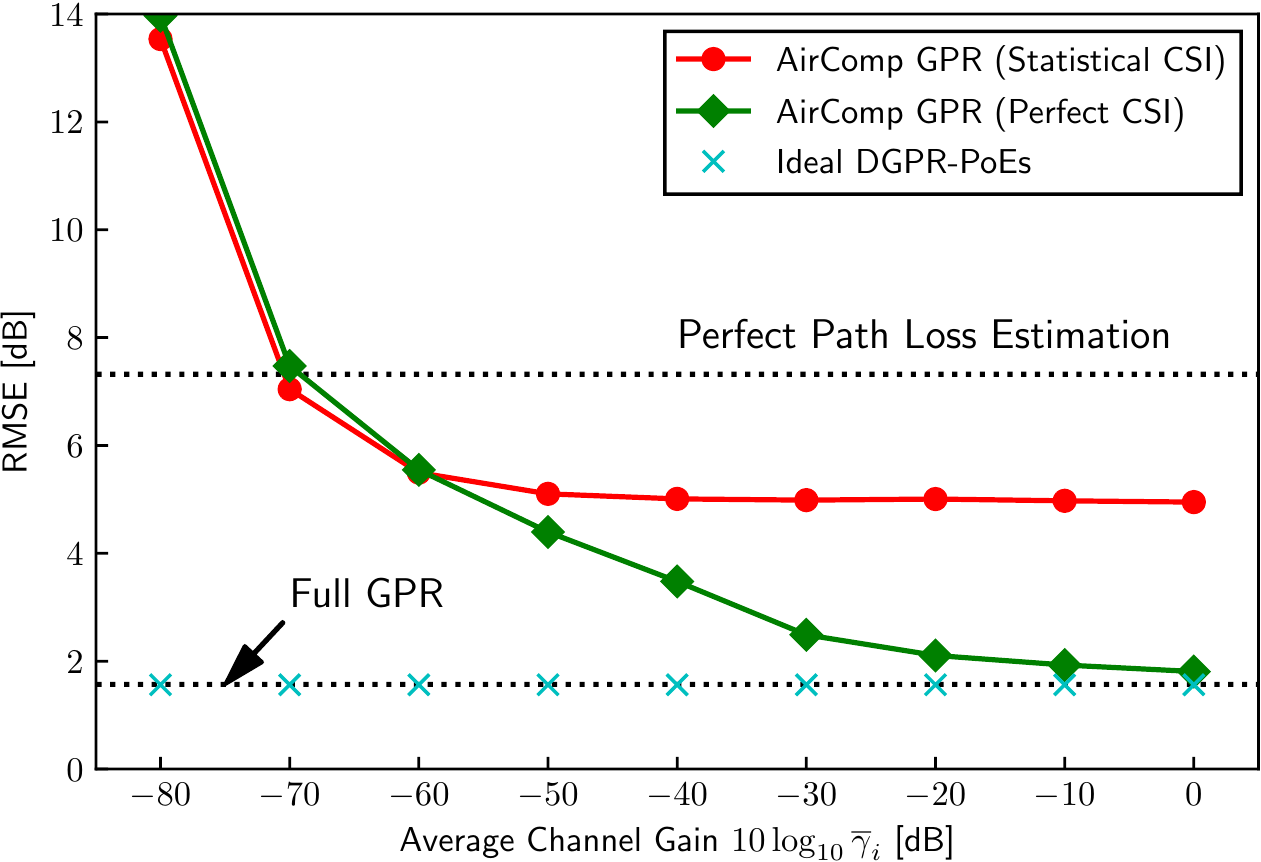}
  \caption{Effects of $\overline{\gamma}_i$ on RMSE.}
  \label{fig:effects-gain}
%   \label{fig:effects-nodes}
\end{figure}

The effect of channel gain on the RMSE is shown in Fig.\ref{fig:effects-gain}.
Regression accuracies of the AirComp-based methods tend to be degraded owing to AWGN. In this case, both perfect and statistical CSI-based methods show better accuracies than the path loss-based method where $\overline{\gamma}_i$ is over -60\,dB.
Further, the gap between the perfect CSI-based method and full GPR was almost zero at $10\log_{10}\overline{\gamma}_i=0$; in contrast, the statistical CSI-based method takes 3.38\,dB.
AirComp GPR with the statistical CSI-based training is affected by statistical amplitude fluctuation, as shown in Eq.\,\eqref{eq:decoder_stat}. Since Nelder-Mead simplex is a deterministic approach, the hyperparameter may not be trained fully; thus, this approach showed this RMSE gap.
\par
\begin{figure}[t]
    \centering
    \includegraphics[width=0.9\linewidth]{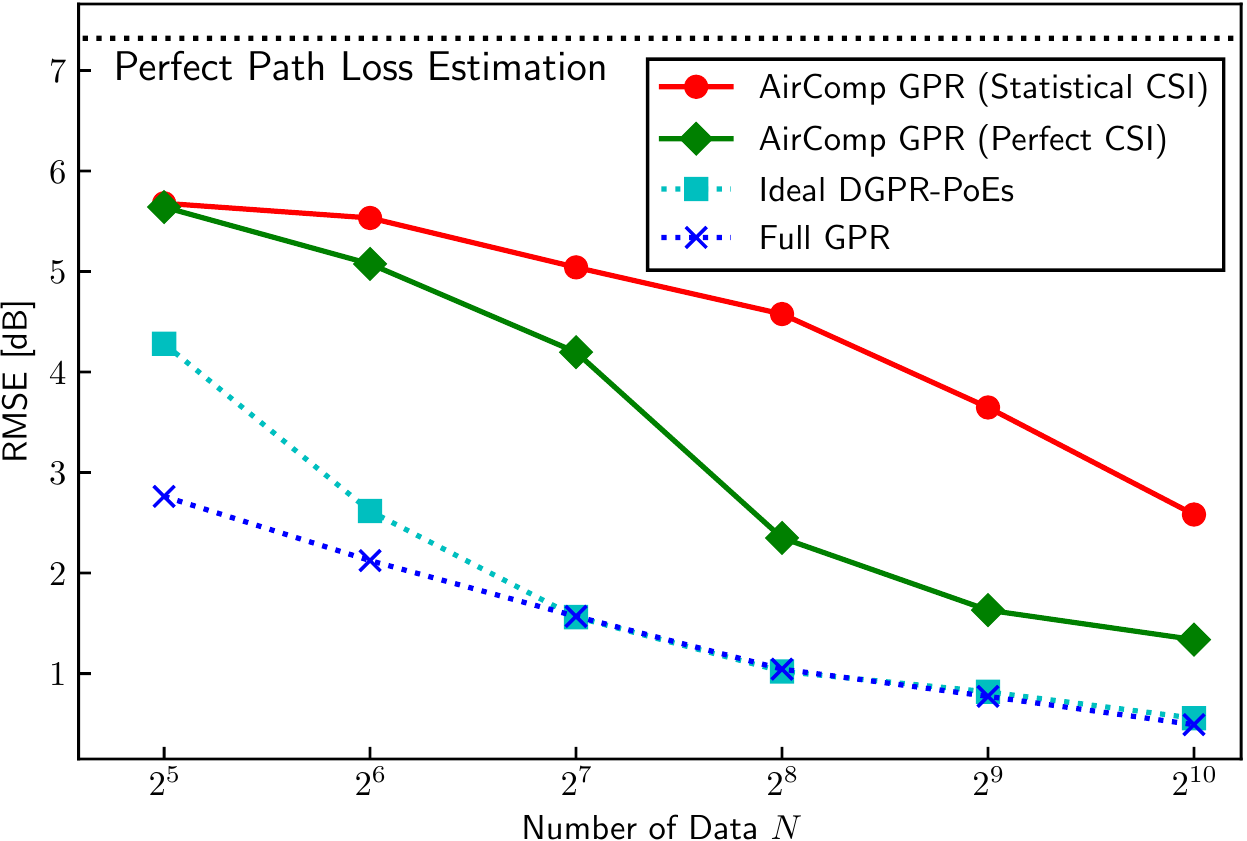}
  \caption{Effects of $N$ on RMSE.}
  \label{fig:effects-n}
%   \label{fig:effects-nodes}
\end{figure}
\begin{figure}[t]
    \centering
    \subfigure[$N=128$.]{\includegraphics[width=0.48\linewidth]{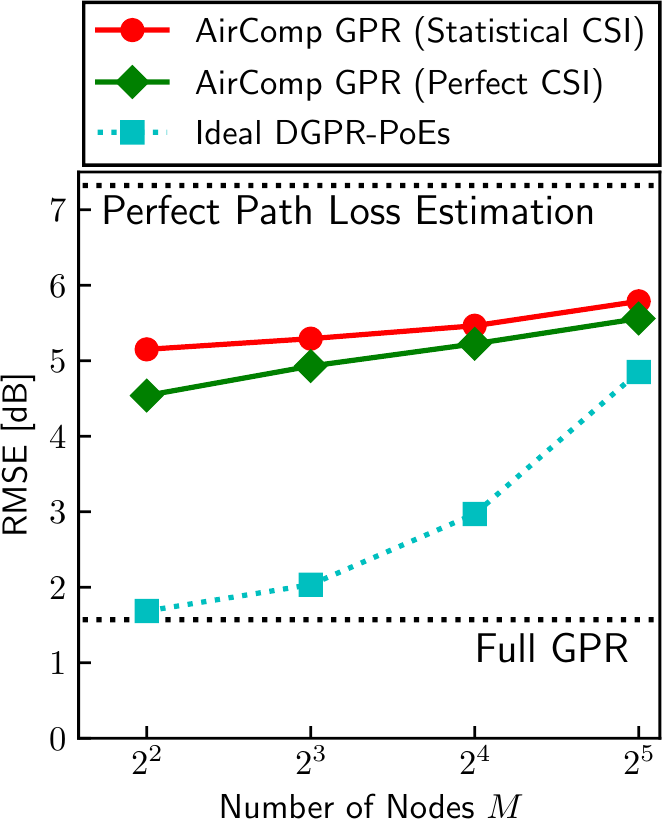}
    \label{subfig:n128}}
    \subfigure[$N=512$.]{\includegraphics[width=0.48\linewidth]{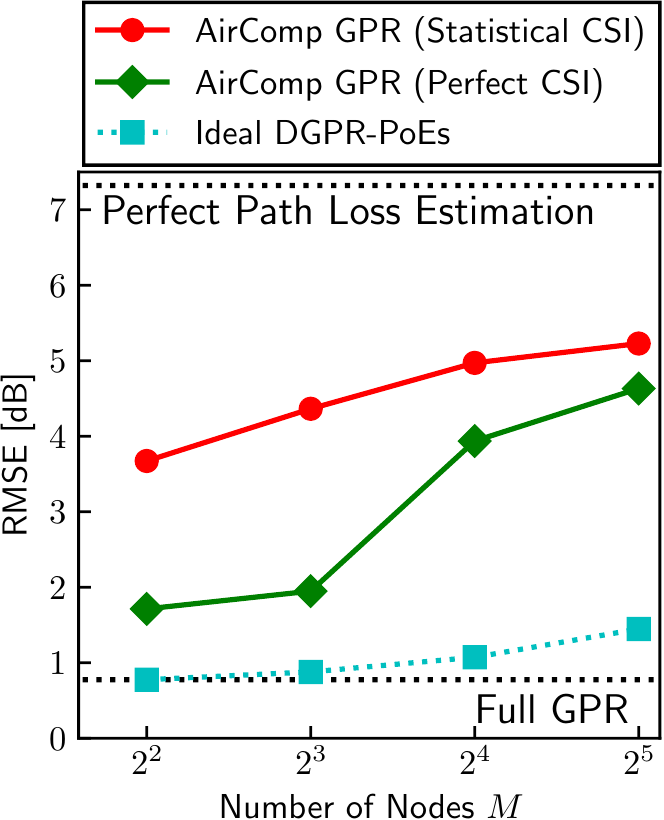}
    \label{subfig:n512}}
  \caption{Effects of $M$ on RMSE.}
  \label{fig:effects-nodes}
\end{figure}

We show effect of the number of data $N$ in Fig.\,\ref{fig:effects-n}.
All GPR-based methods can improve the accuracy performance as $N$ increases.
The gaps between full GPR and AirComp-based methods when $N=2^{10}$ were 0.85\,dB in perfect CSI-based method and 2.09\,dB in statistical CSI-based method.
Finally, effects of the number of nodes $M$ are demonstrated in Fig.\,\ref{fig:effects-nodes}.
Gaps between full GPR and AirComp-based methods increase at many nodes since it divides the likelihood into multiple local pieces.
However, Fig.\,\ref{fig:effects-nodes} reveals that the AirComp-based methods could achieve better accuracy than the path loss-based method in various conditions. 
Comparing perfect CSI-based and statistical CSI-based methods, the gap is small for $N=128$; e.g., 0.04\,dB at $M=2^5$.
\par
In summary, AirComp GPR can construct more accurate radio maps than the perfect path loss estimation in various conditions with communication and computation efficiencies.


%% ノード数の影響
%% ノード数が増えると近似が甘くなるため精度が落ちがち

\section{Conclusion}
We proposed an over-the-air computation-aided distributed GPR scheme, termed AirComp GPR, with both perfect CSI-based and statistical CSI-based training methods.
Our simulation demonstrated that AirComp GPR speeds up the computation time roughly 733x than full GPR when $N=2^{10}$ and $M=16$ while maintaining its communication cost constant regardless of the numbers of nodes and training data.
AirComp GPR will enable low-latency regression analysis over distributed IoT networks.

\bibliographystyle{IEEEbib}
\bibliography{reference}

\end{document}